\documentclass[fleqn,usenatbib]{mnras}
\usepackage{newtxtext,newtxmath}
\usepackage[T1]{fontenc}
\usepackage{graphicx}
\usepackage{siunitx}

\newcommand{\Ixx}{I_{xx}}
\newcommand{\Iyy}{I_{yy}}
\newcommand{\Izz}{I_{zz}}
\newcommand{\MCin}{^{\text{in}}}
\newcommand{\MCout}{^{\text{out}}}
\newcommand{\MCset}{\mathcal{MC}}
\newcommand{\NRE}{\bar{\mathcal{E}}}
\newcommand{\RE}{\mathcal{E}}
\newcommand{\cA}{\mathcal{A}}
\newcommand{\cD}{\mathcal{D}}
\newcommand{\cN}{\mathcal{N}}
\newcommand{\cO}{\mathcal{O}}
\newcommand{\dist}{r}
\newcommand{\fddot}{\ddot{f}}
\newcommand{\fdot}{\dot{f}}
\newcommand{\pmp}{m_p}

\DeclareMathOperator{\median}{median}

\title[Inferring NS properties with CWs]{Inferring neutron star properties with continuous gravitational waves}
\author[Neil Lu, Karl Wette, Susan M. Scott, Andrew Melatos]{
  Neil Lu,$^{1,2}$\thanks{E-mail: neil.lu@anu.edu.au}
  Karl Wette,$^{1,2}$\thanks{E-mail: karl.wette@anu.edu.au}
  Susan M. Scott$^{1,2}$
  and Andrew Melatos$^{1,3}$
  \\
  $^{1}$ ARC Centre of Excellence for Gravitational Wave Discovery (OzGrav), Hawthorn VIC 3122, Australia \\
  $^{2}$ Centre for Gravitational Astrophysics, Australian National University, Canberra ACT 2601, Australia \\
  $^{3}$ School of Physics, University of Melbourne, Parkville VIC 3010, Australia
}

\pubyear{2022}

\begin{document}
\label{firstpage}
\pagerange{\pageref{firstpage}--\pageref{lastpage}}
\maketitle

\begin{abstract}
  Detection of continuous gravitational waves from rapidly-spinning neutron
  stars opens up the possibility of examining their internal physics. We develop
  a framework that leverages a future continuous gravitational wave detection to
  infer a neutron star's moment of inertia, equatorial ellipticity, and the
  component of the magnetic dipole moment perpendicular to its rotation axis. We
  assume that the neutron star loses rotational kinetic energy through both
  gravitational wave and electromagnetic radiation, and that the distance to the
  neutron star can be measured, but do not assume electromagnetic
  pulsations are observable or a particular neutron star equation of state.
  We use the Fisher information matrix and Monte Carlo
  simulations to estimate errors in the inferred parameters, assuming a
  population of gravitational-wave-emitting neutron stars consistent with the
  typical parameter domains of continuous gravitational wave searches. After an
  observation time of one year, the inferred errors for many neutron stars are
  limited chiefly by the error in the distance to the star. The techniques developed here will be useful if continuous
  gravitational waves are detected from a radio, X-ray, or gamma-ray pulsar, or
  else from a compact object with known distance, such as a supernova remnant.
\end{abstract}



\section{Introduction}

The field of gravitational wave astronomy, while relatively new, has the
potential to make exciting contributions to many areas of astrophysics. The
first gravitational wave event, GW150914, was detected in 2015 and was generated
by a binary black hole merger~\citep{Abbott2016:GW150914}. Since this first
detection, the Laser Interferometer Gravitational-Wave Observatory (LIGO)~\citep{LIGO2015:AdvLIG} and
Virgo~\citep{Virg2015:AdVScnIntGrWDt} gravitational wave observatories have made
almost 100 confirmed detections~\citep{Abbott2021:GWTC2}. The KAGRA detector \citep{Akutsu2019:KAGRA} also operated during the later portion of the O3b observing run \citep{Akutusu2020:KAGRA_overview}.

Gravitational wave detection allows for multi-messenger astronomy if an event is
simultaneously observed in both electromagnetic and gravitational wave
bands. This was successfully achieved in 2017 with the detection of GW170817, a
gravitational wave event generated by a binary neutron star merger, which was
also independently observed as the gamma ray burst GRB170817a across the electromagnetic
spectrum~\citep{Abbott2017:GW170817, Abbott2017:GW170817a}. Gravitational wave and multi-messenger
astronomy will likely be the source of many future discoveries as gravitational
wave detector sensitivity increases and more detectors come online.

Gravitational wave astronomy has the potential to advance our understanding of
neutron stars and the physics of matter at extreme densities. Gravitational waves from neutron star
mergers are detectable by current ground-based observatories, but only for a short
time at the end of their life cycle when their stellar structure is tidally
deformed.

Continuous gravitational waves are long-lived, quasi-monochromatic gravitational
waves emitted by isolated spinning neutron stars that are deformed asymmetrically
about their rotation axes. The deformation may be caused by a number of
mechanisms, including the neutron star's magnetic
field~\citep{ZimmSzed1979:GrvWRtPrRBSMAppPl, BonaGour1996:GrvWPlEmMgFIDst},
magnetically-confined mountains~\citep{MelaPayn2005:GrvRdAcMlPMgnCM}, or
electron capture gradients~\citep{UshoEtAl2000:DfrAcNtSCGrvWEm}. The stellar
structure is expected to be in a long-lived stable
equilibrium and continuous gravitational waves (hereafter abbreviated to
``continuous waves'') could provide insights into the ground state of nuclear matter
complementary to observations of binary neutron star mergers.

While continuous waves have not yet been detected, prospects for a first
detection continue to improve with more sensitive gravitational wave
detectors. In addition, the data analysis techniques used to search for
continuous wave signals continue to be refined; for reviews see \citet{Riles2013,
  Riles2017, Tenorio2021}. Searches for continuous waves cover a wide variety of
sources and include: known radio and X-ray
pulsars~\citep{LIGOEtAl2020:GrvCnsEqElMlP, LIGOEtAl2021:DBSpLCnsGrvWEnYPPJ0,
  LIGOEtAl2021:CnsLODGrvEmDRGlPPJ0, LIGOEtAl2022:SCnGrvW20AcMllXPlOLD,
  LIGOEtAl2022:NrSCnLngTrGrvWKPLTOR, LIGOEtAl2022:SrGrvWKPTHrmSTLIObR}, likely
neutron stars in supernova remnants~\citep{LIGOEtAl2021:SrCntGrvWYSpRETObRALV,
  LIGOEtAl2022:SEOLDCntGrvWCsVJSpRm}, and all-sky surveys for undiscovered
neutron stars~\citep{LIGOVirg2021:ASEOLDCntGrvSgUnNtSBS,
  LIGOEtAl2021:AlSCntGrvWIsNtSEOLD, LIGOEtAl2022:AlSrGrvWEmsSBCASpBHLOD,
  CovaEtAl2022:CnsRMnMllNtSBSy}.

In this paper, we study what macroscopic properties of neutron stars might be
inferred using continuous waves. \citet{Sieniawska2021} have previously studied
this question under the assumption that the neutron star loses rotational
kinetic energy purely through gravitational wave radiation. Here we consider a
more general model where energy is also lost through electromagnetic radiation,
assuming the star possesses a dipolar magnetic field. The population dynamics of
neutron stars losing energy through both electromagnetic and gravitational
radiation have previously been studied by \citet{Palomba2005, Knispel2008,
  Wade2012, Cieslar2021, ReedEtAl2021:MdGlNSPplUCnGrvS}. This paper is an initial attempt at studying the
parameter estimation problem for such systems.

This paper is organised as follows. Section~\ref{sec:background} introduces
background information on continuous waves and their
detection. Section~\ref{sec:framework} introduces the theoretical framework used
to infer properties of the neutron star. Section~\ref{sec:MC} discusses how
Monte Carlo simulations are used to estimate the errors of the inferred
properties. Section~\ref{sec:results} presents the results of the Monte Carlo
simulations. Section~\ref{sec:assumptions} considers some of the caveats and
assumptions, and Section~\ref{sec:summary} summarises the results.

\section{Background} \label{sec:background}

This section presents background information relevant to the paper. Subsections \ref{sec:CW_signal} and \ref{sec:error_theory} introduce the basics of the signal model and parameter estimation techniques respectively for continuous wave searches.

\subsection{Continuous wave signal model} \label{sec:CW_signal}

A continuous wave induces a strain $h(t)$ in a gravitational wave detector given
by~\citep{Jaranowski1998}:
\begin{equation}
  h(t) = \sum_{i=1}^4 \cA_i h_i(t; \vec\lambda) \,. \label{eq:hoft}
\end{equation}
The four amplitudes $\{\cA_i\}$ are functions of: the characteristic
strain amplitude $h_0$, the inclination angle $\iota$ of the neutron star's
angular momentum to the line of sight, a polarisation angle $\psi$ which fixes
the principal axes of the two gravitational wave polarisations (``plus'' and
``cross''), and an arbitrary phase $\phi_0$ at a reference time $t_0$.
Additional parameters, represented by $\vec\lambda$ in Eq.~\eqref{eq:hoft},
modify the phase of the signal; they include the star's sky position and, if the
star is in a binary system, its orbital parameters.

The characteristic strain amplitude $h_0$ of a continuous wave signal
is~\citep{Jaranowski1998}:
\begin{gather}
  h_0 = \frac{4\pi^2G}{c^4}\frac{\Izz \epsilon f^2}{\dist} \,, \label{eq:h0}
\end{gather}
where $\dist$ is the
distance to the neutron star, $f$ is the gravitational wave frequency, $G$ is
the gravitational constant, and $c$ is the speed of light.
We model the neutron star as a tri-axial
rotor~\citep{ZimmSzed1979:GrvWRtPrRBSMAppPl} with principal moments of inertia
$(\Ixx, \Iyy, \Izz)$, where the $z$ axis points along the star's symmetry rotation axis; the equatorial ellipticity $\epsilon = |\Ixx - \Iyy| / \Izz$ characterises the
degree of non-axisymmetrical deformation of the star. For a tri-axial
rotor the gravitational wave frequency is conventionally assumed to be twice the star's rotational
frequency~\citep{VanD2005:GrvWSpNnxFPrNtS, Sieniawska2021}.

The radiation of rotational kinetic energy away from the neutron star via
continuous waves, and possibly electromagnetic radiation, causes the star to
spin down. We model the evolution of the gravitational wave frequency as a
second-order Taylor expansion~\citep{Jaranowski1998}:
\begin{gather}
  f(t) = f + \fdot t + \frac{1}{2}\fddot t^2 \, ,  \label{eq:signal_f}
\end{gather}
where $f$ is the gravitational wave frequency and $\fdot$ and $\fddot$ its
first and second time derivatives respectively, where all three parameters are defined
at $t=t_0$. These parameters enter into Eq.~\eqref{eq:hoft} as phase parameters
represented by $\vec\lambda$.

A useful quantification of the spin-down behaviour of a neutron star is its
braking index~\citep{MancEtAl1985:ScMsrPlBrkIn}:
\begin{gather}
  n = \frac{f\fddot}{\fdot^2} \, . \label{eq:braking_index}
\end{gather}
If a neutron star is spinning down purely through the emission of gravitational
waves from a (mass-type) quadrupole moment, as given by Eq.~\eqref{eq:h0}, its
braking index is $n=5$; alternately, if the neutron star spins down only through
electromagnetic radiation, its braking index is
$n=3$~\citep{OstrGunn1969:NtrPlsITh} but cf.
\citep{Mela1997:SpObRtCrrOMgn}. A third possibility, which we do not
consider in this paper, is the emission of gravitational waves from a
current-type quadrupole moment due to $r$-modes~\citep{Ande1998:NClUnsMdRtRltS,
  LindEtAl1998:GrvRdInsHYnNtS}, for which one has $n=7$.

\subsection{Continuous wave parameter estimation}\label{sec:error_theory}
Detection of a continuous wave signal would measure its amplitude and phase to some degree of uncertainty, assuming that the true continuous
wave signal does not deviate appreciably from the model described in
section~\ref{sec:CW_signal}. Bayesian inference is widely regarded as a robust method of
inferring parameters of a signal model given a data-set and assumed priors on the
parameters; for its application to continuous waves see
\citet{DupuWoan2005:ByEstPlPrGrvWD, PitkEtAl2017:NSmCTrSrCntGrvWP}.

As an initial attempt to study the errors in the parameters measured by a
continuous wave detection, we instead adopt a simpler approach using the Fisher
information matrix. While this approach is commonly used \citep{Sieniawska2021,
  Jaranowski1999}, the Fisher information matrix is strictly valid only in the
case of high signal-to-noise ratios (a criterion for which is detailed in
\citealt{Vallisneri2008}), which may not necessarily be the case for a
first continuous wave detection. Further discussion of the weaknesses of
the Fisher information matrix approach, such as the possibility of singular or ill-conditioned Fisher information matrices, are outlined in
\citet{Vallisneri2008}. Notwithstanding these concerns, we use the Fisher
information matrix because of its relative computational simplicity
to arrive at a quantitative picture of parameter inference for continuous wave
signals. We now outline how Fisher information matrices can be used to
approximate the error of the continuous wave parameters.

Data analysis techniques that seek to identify continuous waves often quantify
how closely an observed signal matches a template of possible signals.
An intuitive picture of how the Fisher information matrix works is that
it quantifies the maximal possible ``distance'' in the parameter space that a
true signal could be from the ``nearest'' template. That is, since the template
bank forms a ``grid'' (which may not be uniform) in the parameter space, the maximal error of the parameter
measurements comes from the size of the ``gaps'' in the template bank.

As will become clear in Section~\ref{sec:framework}, we are particularly
interested in estimating errors in the three parameters $f$, $\fdot$, and
$\fddot$ of Eq.~\eqref{eq:signal_f} which govern the gravitational wave frequency
evolution. To construct the Fisher information matrix for these parameters, we
start with phase of the continuous wave signal:
\begin{align}
  \phi_{\text{spin}}(t)
  &= 2 \pi \int_0^{t} f(t') dt'\, , \label{eq:GW phase temp}\\
  &= 2\pi \left[f t + \frac{1}{2}\fdot(t) t^2
    + \frac{1}{6} \fddot(t)t^3 \right] \, , \label{eq:GW phase}
\end{align}
where Eq.~\eqref{eq:signal_f} has been substituted into Eq (\ref{eq:GW phase temp}).  We next compute the
parameter-space
metric~\citep{BalaEtAl1996:GrvWClsBDStMCEsPr,Owen1996:STmGrvWInsBnCTmS} which
quantifies the notion of ``distance'' between the true signal and a template.
It is necessary to first define the time average operator:
\begin{gather}
  \Big \langle x(t) \Big \rangle = \frac{1}{T} \int_{-T/2}^{T/2} x(t) dt \, ,
\end{gather}
where $x(t)$ is an arbitrary function and $T$ is the time span of the
gravitational wave observation. The parameter-space metric over $f$, $\fdot$,
and $\fddot$ is then given by~\citep{BradEtAl1998:SrcPrdSrLI,Prix2007}
\begin{gather} \label{eq:template metric}:
  g_{i j} =
  \Bigg\langle
  \frac{\partial \phi_{\text{spin}}(t)}{\partial f^{(i)}}
  \frac{\partial \phi_{\text{spin}}(t)}{\partial f^{(j)}}
  \Bigg\rangle
  -
  \Bigg \langle
  \frac{\partial \phi_{\text{spin}}(t)}{\partial f^{(i)}}
  \Bigg \rangle
  \Bigg \langle
  \frac{\partial \phi_{\text{spin}}(t)}{\partial f^{(j)}}
  \Bigg \rangle \, ,
\end{gather}
with $i, j \in \{0, 1, 2\}$, $f^{(0)} = f$, $f^{(1)} = \fdot$, and
$f^{(2)} = \fddot$.

The covariance matrix for $f$, $\fdot$, and $\fddot$ is given by the inverse
Fisher information matrix $\Gamma^{ij}$~\citep{Vallisneri2008}, which, in turn, is
defined in terms of the parameter-space metric~\citep{Prix2007}:
\begin{align} \label{eq:Fisher matrix}
  \Sigma(f, \fdot, \fddot) &= \Gamma^{ij} \\
  &= \frac{g^{ij}}{\rho^2} \,.
\end{align}
Here $\rho^2$ is the signal-to-noise ratio assuming an optimal match between the
true signal and the best-fit template.  For observation times of a year or more,
we can assume an expression for $\rho^2$ averaged over $\cos\iota$, $\psi$, and sky
position~\citep{Jaranowski1998,Prix2011}:
\begin{align}
  \rho^2 &= \frac{4}{25}\frac{h_0^2 T}{S_h(f)} \, , \label{eq:rho2_using_h0} \\
         &= \frac{4}{25}\frac{T}{\cD^2} \, , \label{eq:rho2_using_depth}
\end{align}
where $S_h$ is the (single-sided) power spectral density of the strain noise in
the gravitational wave detector, and Eq. \eqref{eq:rho2_using_depth} defines the "sensitivity depth"~\citep{BehnEtAl2015:PstMtUSCnGrvSGlC,
  DreiEtAl2018:FAcSnsEsCntSr}:
\begin{equation}
  \cD = \frac{ \sqrt{S_h(f)} }{ h_0 } \, , \label{eq:sens-depth}
\end{equation}
We assume, again for simplicity, that the
gravitational wave detector network is operational at 100\% duty cycle; in
practice duty cycles of $\gtrsim 70\%$ are achieved for current detectors,
but this is expected to improve over time~\citep{KAGREtAl2020:PrObLclGrvTrALAVK}.
Evaluating Eq.~\eqref{eq:Fisher matrix} using
Eqs.~\eqref{eq:GW phase} and~\eqref{eq:template metric} gives the covariance matrix:
\begin{align}
  \Sigma(f, \fdot, \fddot) &= \frac{ \cD^2 }{ \pi^2 } \begin{pmatrix}
    \frac{ 1875 }{ 16 T^3 } & 0 & -\frac{ 7875 }{ 2 T^5 } \\
    0 & \frac{ 1125 }{ T^5} & 0 \\
    -\frac{ 7875 }{ 2 T^5 } & 0 & \frac{ 157500 }{ T^7 }
  \end{pmatrix} \, . \label{eq:spindown_cov}
\end{align}

Now considering the four amplitude parameters $h_0$, $\cos\iota$, $\psi$, $\phi_0$; only $h_0$ is potentially interesting for inferring neutron star properties as it is a
function of $\Izz$ and $\epsilon$ [Eq.~\eqref{eq:h0}].  The error in the $h_0$ measurement may be derived from the parameter-space metric over the
amplitude parameters $\{\cA_i\}$~\citep{Prix2007}, as outlined in
\citet[Section 3.2]{Prix2011}.
For year-long observations, it is conventional to consider the error in $h_0$
averaged over sky position~\citep[Eq.~122]{Prix2011} and $\psi$:
\begin{align}
  \label{eq:h0_error}
  \sigma(h_0) &= \frac{ a \cD h_0 }{ \sqrt{T} } \frac{ \sqrt{ b + \xi^2 } }{ 1 - \xi^2 } \,, \\
  \xi &\equiv \cos\iota \,, \\
  a &= 2 \sqrt{\frac{6}{301}\left(344 - 43\sqrt{2} - 8\sqrt{86}\right)} \approx 4.08 \,, \\
  b &= \frac{43 \left(8 - 8\sqrt{2} - \sqrt{86}\right)}{43\sqrt{2} + 8\sqrt{86} - 344} \approx 2.59 \,.
\end{align}
Note that Eq.~\eqref{eq:h0_error} becomes infinite at $\xi = \pm 1$, due to a
singularity in the coordinate transform between $\{\cA_i\}$ and
$\{h_0, \xi, \psi, \phi_0\}$; for this reason Eq.~\eqref{eq:h0_error} cannot be
analytically averaged over $\xi$ with a prior range that includes $\pm 1$.

\section{Parameter estimation framework}  \label{sec:framework}
This section develops a framework for inferring three neutron star properties:
its principal moment of inertia ($\Izz$), its ellipticity
($\epsilon$), and the component of the magnetic dipole moment perpendicular to
its rotation axis ($\pmp$, hereafter abbreviated to ``perpendicular magnetic
moment''). It assumes that the neutron star is losing rotational kinetic energy
(and hence spinning down) through both magnetic dipole radiation and
gravitational wave (mass-type) quadrupolar radiation, and that no other mechanisms
dissipate energy from the neutron star. This framework relies on the detection of
a continuous wave signal to measure the frequency and spin-down parameters
($f$, $\fdot$, and $\fddot$), and the characteristic strain amplitude
($h_0$). It also assumes that a measurement of the distance to the neutron star
($\dist$) is available.

Balancing the spin-down power with the luminosity of electromagnetic
and gravitational radiation gives:
\begin{gather}
  \left(\frac{dE}{dt}\right)_{\text{EM}} + \left(\frac{dE}{dt}\right)_{\text{GW}}
  =
  -\left(\frac{dE}{dt}\right)_{\text{rot}} \, . \label{eq:energy_balance}
\end{gather}
The ellipticity of a neutron star is conventionally assumed to be relatively small~\citep{Sieniawska2021}, so the star
is very close to spherical. The rotational kinetic energy of the star is then
taken to be that of a rotating sphere~\citep{WettEtAl2008:SrGrvWvCssLI}:
\begin{gather}
  \Bigg(\frac{dE}{dt}\Bigg)_{\text{rot}}
  =
  \pi^2 \Izz f \fdot \, . \label{eq:rot_energy}
\end{gather}
The luminosity of a rotating magnetic dipole
is~\citep{OstrGunn1969:NtrPlsITh, CondRans2016:EssRdAst}
\begin{gather}
  \left(\frac{dE}{dt}\right)_{\text{EM}}
  =
  \frac{2\pmp^2}{3c^3 \mu_0} ( \pi f )^4 \, , \label{eq:EM_energy}
\end{gather}
where $\mu_0$ is the vacuum permeability. Note that this is given in terms of the gravitational wave frequency which is twice the rotational frequency as discussed in Section \ref{sec:CW_signal}. The
gravitational wave luminosity of a (mass-type) quadrupole
is~\citep{OstrGunn1969:NtrPlsITh, BlanEtAl2001:GrvRdThLgPrp}
\begin{gather}
  \Bigg( \frac{dE}{dt}\Bigg)_{\text{GW}}
  =
  \frac{32G}{5c^5}\Izz^2 \epsilon^2 (\pi f)^6 \, . \ \label{eq:GW_energy}
\end{gather}
In order to simplify the expressions, we introduce the constants
\begin{gather}
  K_{\text{EM}} = \frac{2 \pi^2}{3c^3 \mu_0} \, , \qquad
  K_{\text{GW}} = \frac{32 G \pi^4}{5c^5} \, .
\end{gather}
We then substitute Eqs.~\eqref{eq:rot_energy}~--~\eqref{eq:GW_energy} into
Eq.~\eqref{eq:energy_balance} and rearrange to get:
\begin{gather}
  \fdot = -\frac{K_{\text{EM}} \pmp^2 f^3}{\Izz}
  - K_{\text{GW}} \Izz \epsilon^2 f^5 \, . \label{eq:simul_fd}
\end{gather}
Differentiating Eq.~\eqref{eq:simul_fd} with respect to time gives:
\begin{gather}
  \fddot = -\frac{3 K_{\text{EM}} \pmp^2 f^2 \fdot}{\Izz}
  - 5 K_{\text{GW}} \Izz \epsilon^2 f^4 \fdot \, . \label{eq:simul_fdd}
\end{gather}
Given that $\fddot$ is measured as a separate parameter of the continuous wave
signal model [Eq.~\eqref{eq:signal_f}], Eq.~\eqref{eq:simul_fdd} provides an
additional constraint independent of Eq.~\eqref{eq:simul_fd}.

Equations~\eqref{eq:simul_fd} and~\eqref{eq:simul_fdd} depend on three unknowns:
$\Izz$, $\epsilon$, and $\pmp$. With the addition Equation.~\eqref{eq:h0}
which also depends on $\Izz$ and $\epsilon$, we have three equations
constraining the same three unknowns which may now be solved for:
\begin{align}
  \Izz
  &= \frac{ K_{\text{GW}} c^8 \dist^2 h_0^2 f }{ 8\pi^4 G^2 \fdot ( 3 - n )}
    \, , \label{eq:Izz} \\
  \epsilon
  &= \frac{ 2\pi^2 G \fdot ( 3 - n ) }{ K_{\text{GW}}c^4 \dist h_0 f^3 }
    \, , \label{eq:epsilon} \\
  \pmp
  &= \frac{ c^4 \dist h_0 }{ 4\pi^2 G f }
    \sqrt{ \frac{ K_{\text{GW}} ( n - 5 ) }{ K_{\text{EM}}(3 - n ) } }
    \, , \label{eq:mp}
\end{align}
where $n$ is the braking index of Eq.~\eqref{eq:braking_index}. It is also possible to solve for the mass quadrupole moment of the neutron star \citep{Owen2005:MxElDfrCmSEEqtS}:
\begin{equation}
    Q_{22} = \sqrt{\frac{15}{8 \pi}} \Izz \epsilon
    = \sqrt{ \frac{15}{128 \pi^5} } \frac{c^4 \dist h_0}{G f^2} \, . \label{eq:Q22}
\end{equation}
Note that Eq.~\eqref{eq:Q22} has the nice property of being independent of $n$. Nevertheless, we choose to consider $\Izz$ and $\epsilon$ separately in this work to distinguish neutron star properties with relatively small ($\Izz$) and large ($\epsilon$) prior uncertainties (see Section~\ref{sec:MC_params}).

Equations~\eqref{eq:Izz}~--~\eqref{eq:mp} remain valid provided that $3 < n < 5$,
which is consistent with the power balance assumed by
Eq.~\eqref{eq:energy_balance}.  As discussed in Section~\ref{sec:CW_signal},
braking indices of 3 or 5 correspond to pure electromagnetic or gravitational
wave radiation respectively; in either case, Eqs.~\eqref{eq:Izz}~--~\eqref{eq:mp}
are no longer applicable. A combination of electromagnetic and gravitational
wave radiation yields a braking index between 3 and 5; the loss of kinetic
rotational energy through \emph{both} gravitational wave \emph{and}
electromagnetic radiation is therefore a fundamental requirement of the
framework outlined here.  \citet{Sieniawska2021} show that, for a neutron star only emitting continuous waves and not electromagnetic radiation, degeneracies prevent direct inference of the neutron star
properties without a measurement of $\dist$, which is unlikely to be measurable
without an electromagnetic counterpart.

Few other techniques exist to directly measure $\Izz$. \citet{Damour1988}
propose a method which requires higher-order relativistic corrections to the
periastron advance to be measurable, which is possible only for very
rapidly-spinning binary pulsars. To date the method has only been applicable to
the double pulsar system PSR~J0737$-$3039~\citep{Bejger2005,
  WorlEtAl2008:NclCnsMmInNtS, SteiEtAl2015:UNtSObsDtCThMITDfr,
  MiaoEtAl2022:MmInPJ07LIGNI}. Note that~\citet{MiaoEtAl2022:MmInPJ07LIGNI} assumes
a neutron star equation of state, whereas the framework derived here does not.
Other methods rely on separate measurements of
the neutron star mass and radius through either electromagnetic observations
and/or detection of gravitational waves from binary neutron star
mergers~\citep{SteiEtAl2015:UNtSObsDtCThMITDfr, MiaoEtAl2022:MmInPJ07LIGNI}.  It
is difficult, however, to measure both properties simultaneously for the one
same neutron star~\citep{MillEtAl2019:PJ0MRdNDImpPrpNtSM}.  No method exists for
directly measuring $\epsilon$ other than through a continuous wave
detection. While $\pmp$ (or equivalently the surface magnetic field strength $B$)
is routinely inferred by assuming pure magnetic dipole radiation from known
pulsars~\citep{Kram2005:Pls}, a measurement of $\pmp$ from a mixed
electromagnetic/gravitational wave pulsar would be of interest as it would provide an independent verification of the existing measurements or provide insight into neutron stars with
different energy loss mechanisms.

The errors in the inferred neutron star properties ($\Izz$, $\epsilon$, $\pmp$) has the following dependencies:
\begin{itemize}

\item The errors of the inferred properties ($\Delta \Izz$, $\Delta \epsilon$, and $\Delta \pmp$) depend on $\Delta f$,
  $\Delta \fdot$, $\Delta \fddot$, $\Delta h_0$, and $\Delta \dist$ [Eqs.~\eqref{eq:Izz}~--~\eqref{eq:mp}];

\item The errors of the spindown parameters ($\Delta f$, $\Delta \fdot$, $\Delta \fddot$) depend on $T$ and $\cD$
  [Eq.~\eqref{eq:rho2_using_depth}];

\item The error $\Delta h_0$ depends on $T$, $\cD$, and $h_0$ [or equivalently $S_h$;
  Eq.~\eqref{eq:sens-depth}] and $\xi$ [Eq.~\eqref{eq:h0_error}];

\item The error $\Delta \dist$ is independent of the other parameters.

\end{itemize}
Therefore, we see that the errors in $\Izz$, $\epsilon$, $\pmp$ depend entirely on: the
observation time ($T$); the strength of the continuous wave signal relative to
the detector noise ($\cD$, $h_0$); the ratio of gravitational wave
``plus'' and ``cross'' polarisations ($\xi$); and the uncertainty in the distance to the star ($\Delta \dist$).

An estimate of the relative errors in $\Izz$, $\epsilon$, and $\pmp$ and their
dependence on the parameters $\Lambda = \{ f, \fdot, \fddot, h_0, \dist \}$ may be
arrived at through differential error
analysis~\citep{BenkEtAl2018:EPrpCMAApAdDsdCn}:
\begin{equation}
  \label{eq:rel-err-dea}
  \frac{ \sigma(\Izz)^2 }{ \Izz^2 }
  =
  \frac{1}{ \Izz^2 } \sum_{x, y \in \Lambda}
  \left(\frac{ \partial \Izz }{ \partial x }\right)
  \left(\frac{ \partial \Izz }{ \partial y }\right)
  \begin{cases}
    \sigma(x)^2 & x = y \,, \\
    \Sigma(x, y) & x \ne y \,,
  \end{cases}
\end{equation}
and similarly for $\sigma(\epsilon)^2 / \epsilon^2$ and
$\sigma(\pmp)^2 / \pmp^2$; where $\sigma(x)$ is the standard deviation in the quantity $x$ and $\Sigma(x,y)$ is the covariance between the quantities $x$ and $y$. This analysis yields, to third order in $1/T$:
\begin{align}
  \label{eq:Izz-rel-err-dea}
  \frac{ \sigma(\Izz)^2 }{ \Izz^2 }
  &= \frac{ 4 \sigma(\dist)^2 }{ \dist^2 } + \frac{ 4 \sigma(h_0)^2 }{ h_0^2 }
    + \frac{16875 \cD^2}{16 \pi ^2 f^2 (n-3)^2 T^3} \,, \\
  \label{eq:eps-rel-err-dea}
  \frac{ \sigma(\epsilon)^2 }{ \epsilon^2 }
  &= \frac{ \sigma(\dist)^2 }{ \dist^2 } + \frac{ \sigma(h_0)^2 }{ h_0^2 }
    + \frac{1875 \cD^2 (9 - 2n)^2}{16 \pi ^2 f^2 (n-3)^2 T^3} \,, \\
  \label{eq:mp-rel-err-dea}
  \frac{ \sigma(\pmp)^2 }{ \pmp^2 }
  &= \frac{ \sigma(\dist)^2 }{ \dist^2 } + \frac{ \sigma(h_0)^2 }{ h_0^2 }
  + \frac{1875 \cD^2 (n^2 - 9n + 15)^2}{16 \pi ^2 f^2 (n-5)^2 (n-3)^2 T^3} \, .
\end{align}
The leading-order terms of the relative errors in $\Izz$, $\epsilon$, and $\pmp$
are the relative errors in $h_0$ and $\dist$. Note that $\sigma (\dist) / \dist$ is
independent of $T$, $\sigma(h_0) / h_0$ scales with $T^{-1/2}$
[Eq.~\eqref{eq:h0_error}], and the remaining terms in
Eqs.~\eqref{eq:Izz-rel-err-dea}~--~\eqref{eq:mp-rel-err-dea} scale with $T^{-3}$
or smaller. Since the distance error $\sigma(\dist) / \dist$ is assumed to be constant, in the limit of $T \to \infty$ the relative errors asymptote to:
\begin{gather}
    \lim_{T\to \infty}\frac{\sigma(\Izz)}{\Izz} = \frac{2\sigma(\dist)}{\dist} \label{eq:Izz_limiting_err} \\
    \lim_{T\to \infty}\frac{\sigma(\epsilon)}{\epsilon} = \frac{\sigma(\dist)}{\dist} \label{eq:epsilon_limiting_err} \\
    \lim_{T\to \infty}\frac{\sigma(\pmp)}{\pmp} = \frac{\sigma(\dist)}{\dist} \label{eq:mp_limiting_err}
\end{gather}
The asymptotic error in $\Izz$ is twice that of the other properties because the relationship between $\Izz$ and $\dist$ is $\Izz \propto \dist^2$ [Eq.~\eqref{eq:Izz}] whereas for the other two parameters it is $\epsilon \propto \dist^{-1}$ and $\pmp\propto \dist$ [Eqs.~\eqref{eq:epsilon} -~\eqref{eq:mp}].

\section{Monte Carlo simulations} \label{sec:MC}

The framework presented in Section~\ref{sec:framework} shows that it is possible to infer three neutron star properties using a continuous waves detection. In this section, we describe how Monte Carlo simulations were used to quantify to what accuracy these properties may be inferred with a detection.

The inference relies on five parameters ($f$, $\dot{f}$, $n$, $h_0$, $\dist$) [Eqs. \eqref{eq:Izz}~--~\eqref{eq:mp}] and the errors of the inference depend on four additional parameters ($T$, $\cD$, $\xi$, $\Delta \dist$) as well as $h_0$. In our simulations we choose to input values of $I_{zz}$ instead of $h_0$ through rearrangement of Eq~\eqref{eq:h0}. Results that directly depend on $h_0$ can be viewed as an optimistic or pessimistic case for the continuous wave signal detectability. In comparison, choices of $I_{zz}$ relate only to the neutron star’s internal physics. While larger values of $I_{zz}$ implicitly lead to
a louder continuous wave signal, this also depends on the other neutron star parameters so does not relate as directly to the signal detectability.

The signal from a neutron star is simulated as the set of input values for the nine parameters ($f\MCin$, $\dot{f}\MCin$, $n\MCin$, $I_{zz}\MCin$, $\dist\MCin$, $T\MCin$, $\cD\MCin$, $\xi\MCin$, $\Delta \dist \MCin$). The properties of the neutron star emitter $(\Izz\MCin, \epsilon\MCin, \pmp\MCin)$ can then be calculated using Eqs. \eqref{eq:Izz}--\eqref{eq:mp}.
Measurement errors $(\delta f, \delta \fddot, \delta \fddot)$ in the
simulated $f\MCin, \fdot\MCin,$ and $\ddot{f}\MCin$ (via $n\MCin$) are drawn from a multivariate
normal distribution with covariance matrix given by Eq.~\eqref{eq:Fisher
  matrix}; the measurement error $\delta h_0$ in $h_0\MCin$ is drawn from a
normal distribution with standard deviation given by Eq.~\eqref{eq:h0_error}.
All other covariances between the parameters
$(f\MCin, \fdot\MCin, \fddot\MCin, h_0\MCin)$ are assumed to be zero.  The
measured parameters of the continuous wave signal are then
\begin{equation}
  \label{eq:MC-output-parameters}
  \begin{aligned}
    f\MCout
    &= f\MCin + \delta f \,,
    & \fdot\MCout
    &= \fdot\MCin + \delta \fdot \,, \\
    \fddot\MCout
    &= \fddot\MCin + \delta \fddot \,,
    & h_0\MCout
    &= h_0\MCin + \delta h_0 \,.
  \end{aligned}
\end{equation}
Substitution of $(f\MCout, \fdot\MCout, \fddot\MCout, h_0\MCout)$ into
Eqs.~\eqref{eq:h0},~\eqref{eq:braking_index} and~\eqref{eq:Izz}~--~\eqref{eq:mp}
gives the inferred neutron star properties
$(\Izz\MCout, \epsilon\MCout, \pmp\MCout)$, which may then be compared to
$(\Izz\MCin, \epsilon\MCin, \pmp\MCin)$.  We repeat this process for $10^6$ samples.

Below we describe the Monte Carlo procedure in further detail.

\subsection{Choice of input parameters} \label{sec:MC_params}
Nine variables control the outputs of the Monte Carlo simulations: $f$, $\dot{f}$, $n$, $I_{zz}$, $\dist$, $T$, $\cD$, $\xi$, $\Delta \dist$. We consider an observation time in the range of $T = 0.5 - 4$ years.  One can expect gravitational wave detector observing runs to last at least a year~\citep{KAGREtAl2020:PrObLclGrvTrALAVK}. A continuous wave signal detected
in a year-long observing run may then be followed up in future and/or archival data.

The neutron star distance $\dist\MCin$ is fixed to $\SI{1}{kpc}$ for
simplicity. Such a distance is within the range where all-sky continuous wave
surveys are sensitive to neutron stars with ellipticities $\epsilon \gtrsim 10^{-6}$ and
emitting at frequencies
$f \gtrsim \SI{100}{Hz}$~\citep{LIGOEtAl2021:AlSCntGrvWIsNtSEOLD}. Given that the
neutron star properties depend on the product $\dist h_0$
[Eqs.~\eqref{eq:Izz}~--~\eqref{eq:mp}], a choice of a smaller (larger) distance
would be equivalent to simulating a larger (smaller) $h_0$.  The
fractional uncertainty in $\dist$ is chosen to be
$\sigma(\dist) / \dist = 20\%$. While radio pulsar
distances (inferred through dispersion measures)
exhibit appreciable variety and are susceptible to biases~\citep{Verbiest2012},
a typical measurement uncertainty of $\sim 20\%$ is not
unreasonable~\citep{Taylor1993, Yao2017}, and indeed is expected to be readily
achievable with next-generation radio telescopes~\citep{Smits2011}.

We explore a range of sensitivity depths $\cD = 30 -
\SI{150}{Hz^{-1/2}}$. The lower end of the range is consistent with the sensitivities
typical of all-sky continuous wave surveys for isolated neutron
stars~\citep[Table~I]{DreiEtAl2018:FAcSnsEsCntSr}; given the wide parameter
space of $f$, $\dot{f}$, and sky position these searches must
cover, their sensitivities are typically lower than targeted continuous wave searches. The upper range is a conservative choice for searches targeting known pulsars~\citep[Table~V]{DreiEtAl2018:FAcSnsEsCntSr}; these
searches cover a much smaller parameter space around the pulsar, and can afford
the computational cost of performing an optimal matched filter analysis to
maximise sensitivity. The range in $\cD$ represents two possible scenarios for a
first continuous wave detection. A continuous wave candidate initially found in
an all-sky survey (with $\cD \sim \SI{30}{Hz^{-1/2}}$) would be followed up with
more sensitive analyses, increasing its signal-to-noise significantly and
yielding a strongly-detected signal. On the other hand, given that searches for
continuous waves from known pulsars already employ the most sensitive methods
(and hence have $\cD \gtrsim \SI{150}{Hz^{-1/2}}$), any signal may initially
only be marginally detectable until more sensitive data becomes available.

We draw the moment of inertia from the widely accepted range for
neutron stars of
$\Izz\MCin \in [1, 3]{\times}10^{38} \, \si{kg.m^2}$~\citep{MolnOstg1985:ClcMMmInrNtSt,
  Bejger2005, WorlEtAl2008:NclCnsMmInNtS, KramEtAl2021:StrGrvTsDbPl,
  MiaoEtAl2022:MmInPJ07LIGNI}.
Ranges for $\epsilon$ and $\pmp$ are less well constrained; estimates
for $\epsilon$ range from $\sim 10^{-11}$~\citep{BonaGour1996:GrvWPlEmMgFIDst}
to $\sim 10^{-4}$~\citep{Owen2005:MxElDfrCmSEEqtS}.  Based on observations of
radio pulsars and magnetars, the surface magnetic field strength
$B = \pmp / R^3$ (where $R$ is the neutron star radius) may range from
$\sim 10^{8}$ to $\sim
10^{15}$~Gauss~\citep{Reis2001:MgnFlNtStOvr}. Certain values of
$(\epsilon, \pmp)$ drawn from these ranges represent neutron stars which spin
down within timescales of seconds to days, which would be impossible to detect as
continuous wave sources. To exclude such regions of the $\epsilon$--$\pmp$
space, we instead draw values of $f\MCin$ and $\fdot\MCin$ from ranges which are
typical of parameter spaces for all-sky continuous wave
surveys~\citep{LIGOEtAl2021:AlSCntGrvWIsNtSEOLD}:
\begin{gather}
  f\MCin \in [50, 2000]~\si{Hz} \,,
  \qquad
  \fdot\MCin \in [-10^{-8}, -10^{-12}]~\si{Hz.s^{-1}} \,.
\end{gather}
A braking index is also drawn from $n\MCin \in (3, 5)$ which is used to compute
$\fddot\MCin$ via Eq.~\eqref{eq:braking_index}.

Having fixed $\dist\MCin$ and chosen $\{\Izz\MCin$, $f\MCin$, $\fdot\MCin$, $\fddot\MCin\}$, we compute $h_0\MCin$ by rearranging Eq. \eqref{eq:Izz}, then calculate $\epsilon\MCin$ and $\pmp\MCin$ via Eqs. \eqref{eq:epsilon} and \eqref{eq:mp}. A choice of $\cD$
then fixes $S_h$ via Eq.~\eqref{eq:sens-depth}. In this paper,   we do not assume a
specific gravitational wave detector configuration (e.g.\ by setting $S_h$ to
the noise power spectral density of a current or future detector). Instead, we assume
that the sensitivity to continuous waves is calibrated by $\dist$ (the
distances which we could detect signals) and $\cD$ (how deep can the data analysis method
dig into the data to extract weak signals). More sensitive gravitational wave detectors will increase the distances $\dist$ at which continuous wave signals may be detected, while improved data analysis
methods will increase our sensitivity to signals, allowing $\cD$ to increase.

There is a strict range for the cosine of the inclination angle $|\xi| \le 1$. As noted
in Section~\ref{sec:error_theory}, however, the error in $h_0$
[Eq.~\eqref{eq:h0_error}] becomes infinite at $|\xi| = 1$ due to a coordinate
singularity. This is a limitation of the analytic Fisher information matrix
approach to error estimation adopted in this paper. That said, the likelihood of
sampling a value of $|\xi| \approx 1$ is negligible.  The use of median and
percentile differences to compare input and output parameters
(Section~\ref{sec:norm-relat-errors}) also guards against degraded Monte Carlo
samples where $|\xi|$ approaches 1. An alternative approach would have been to
assume a particular inclination angle, e.g.\
$\xi = 0$~\citep[cf.][]{Sieniawska2021}.

We do not include sky position in the covariance matrix of Eq.~\eqref{eq:spindown_cov}, as we expect that errors in sky position do not contribute to errors in $(f, \fdot, \fddot)$ when the continuous wave signal is observed for a year or more. We confirmed this assumption with a separate set of Monte Carlo simulations: after extending Eq.~\eqref{eq:spindown_cov} to include sky position, the distribution of errors $(\delta f, \delta \fddot, \delta \fddot)$ were unchanged. Including sky position in the Monte Carlo simulations presented in this work would therefore have had little impact on the results presented in Section~\ref{sec:results}.

\subsection{Computation of output parameters} \label{sec:comp-outp-param}

Having selected the input parameters, output parameters
$(f\MCout, \fdot\MCout, \fddot\MCout, h_0\MCout)$ are computed via
Eq.~\eqref{eq:MC-output-parameters}. An output braking index $n\MCout$ may then
be computed via Eq.~\eqref{eq:braking_index}.

Computation of $(\Izz\MCout, \epsilon\MCout, \pmp\MCout)$ requires
$3 < n\MCout < 5$; this is not guaranteed and can be violated if $n\MCin \approx 3$ or
$n\MCin \approx 5$, and the errors in $\Delta f$, $\Delta \fdot$, and/or $\Delta \fddot$
are also large. Where $3 < n\MCout < 5$ is not satisfied, the Monte Carlo
sample is simply discarded.  At shorter $T$ ($\lessapprox 0.5$ years),
a sizeable fraction ($\gtrapprox 80\%$)
of the samples must be discarded. This fraction decreases with longer $T$, and often
becomes a negligible effect ($\lessapprox 1\%$) once $T \gtrsim 1$~year, but depends
on the exact parameters of the simulation. While this limitation may
impede inference of the properties of a neutron star which is emitting almost
purely electromagnetic or gravitational radiation ($n\MCin \approx 3$ or $\approx 5$
respectively), it is unlikely to be an impediment where an appreciable fraction
of the star's rotational kinetic energy is radiated through both mechanisms.

\subsection{Comparison of inputs and outputs} \label{sec:norm-relat-errors}

The Monte Carlo simulations described above result in pairs
of input and output neutron star properties
$(\Izz\MCin, \Izz\MCout) \in \MCset^{\Izz}$,
$(\epsilon\MCin, \epsilon\MCout) \in \MCset^{\epsilon}$, and
$(\pmp\MCin, \pmp\MCout) \in \MCset^{\pmp}$, where $\MCset$ denotes the results
of the simulations for a particular property.  We quantify the agreement between
input and output properties using the median relative error over each set:
\begin{equation}
  \label{eq:median-relative-error}
  \RE(\Izz) \equiv \median \Bigg\{\,
  \frac{ | \Izz\MCout - \Izz\MCin | }{ \Izz\MCin }
  \,\Bigg|\, (\Izz\MCin, \Izz\MCout) \in \MCset^{\Izz} \,\Bigg\} \,,
\end{equation}
and similarly for $\RE(\epsilon)$ and $\RE(\pmp)$.  From the differential error
analysis of Eq.~\eqref{eq:Izz-rel-err-dea}~--~\eqref{eq:mp-rel-err-dea} it is
expected that, as $T$ increases, $\RE$ will asymptote to a value determine by the
error in the distance $\dist$.  We therefore define normalised relative errors
which asymptote to unity in the limit of $T \to \infty$:
\begin{gather}
  \label{eq:norm-median-relative-error}
  \NRE(\Izz) = \frac{ \RE(\Izz) }{ 2 \RE(\dist) } \\
  \NRE(\epsilon) = \frac{ \RE(\epsilon) }{ \RE(\dist) } \\
  \NRE(\pmp) = \frac{ \RE(\pmp) }{ \RE(\dist) }
  \,,
\end{gather}
where $\RE(\dist)$ is the median error for $\dist$.
Note that $\RE(\Izz)$ is normalised by $2 \RE(\dist)$ due to the quadratic
dependency of $\Izz$ on $\dist$ [Eq. \eqref{eq:norm-median-relative-error}]; see Section~\ref{sec:framework} and Eqs.~\eqref{eq:Izz} and~\eqref{eq:Izz_limiting_err}.
We have assumed
(Section~\ref{sec:MC_params}) a relative error in $\dist$ of 20\%, i.e.\ samples
of $(\dist\MCout - \dist\MCin) / \dist\MCin$ are drawn from a normal
distribution $\cN(0, 0.2)$ with mean zero and standard deviation $0.2$.
Drawing $\sim 10^8$ samples from this distribution gives:
\begin{align}
  \RE(\dist)
  &= \median \Bigg\{\,
    \frac{ | \dist\MCout - \dist\MCin | }{ \dist\MCin }
    \,\Bigg|\, \frac{ \dist\MCout - \dist\MCin }{ \dist\MCin } \sim \cN(0, 0.2) \,\Bigg\} \\
  &\approx 0.135 \,.
\end{align}
We therefore expect $\RE(\Izz)$ to asymptote to $\sim 27\%$, and $\RE(\epsilon)$ and
$\RE(\pmp)$ to asymptote to $\sim 14\%$, at sufficiently large $T$.

\section{Results} \label{sec:results}

\begin{figure}
  \centering
  \includegraphics[width=\columnwidth]{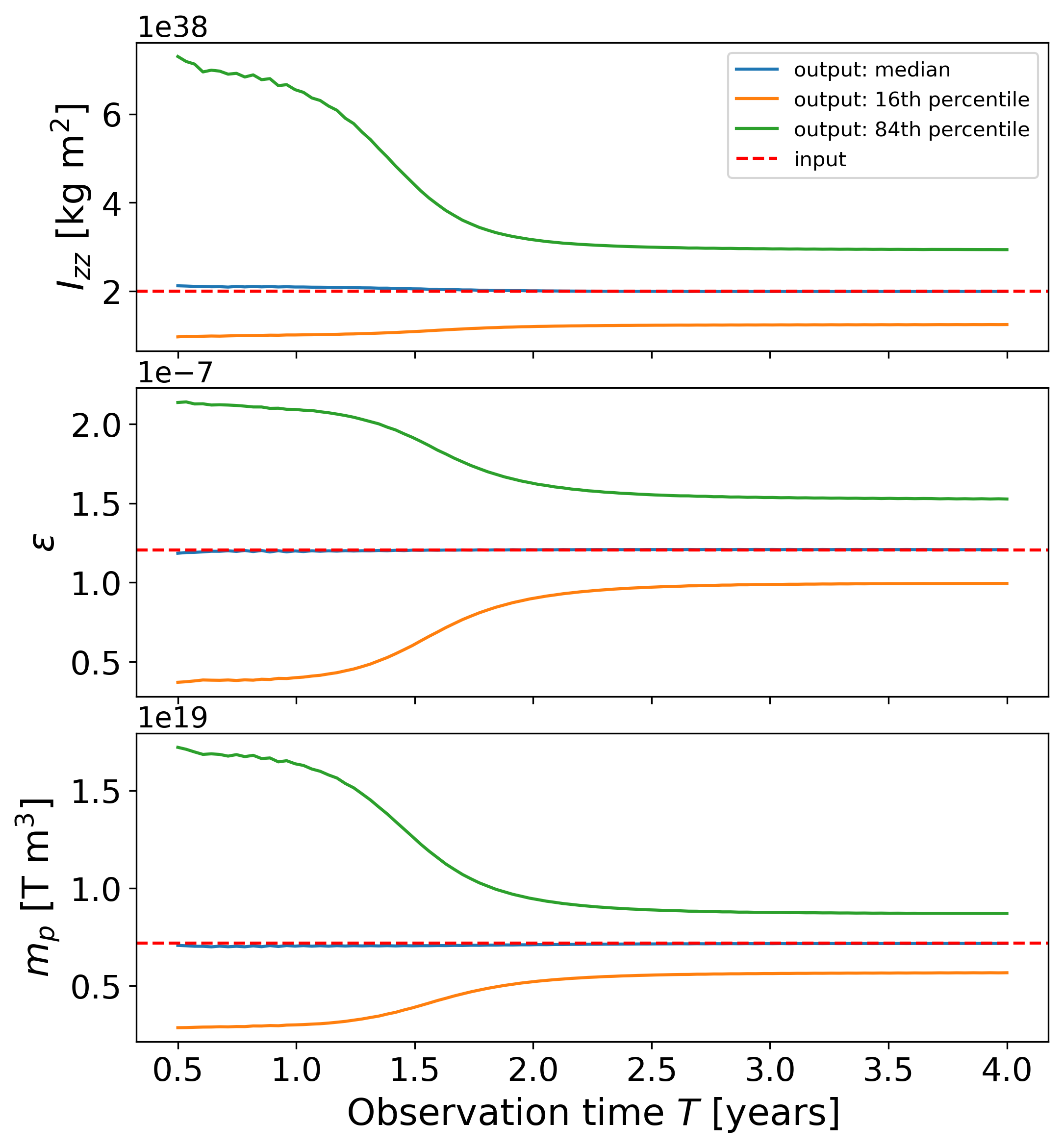}
  \caption{
    Convergence of $(\Izz\MCout, \epsilon\MCout, \pmp\MCout)$ to
    $(\Izz\MCin, \epsilon\MCin, \pmp\MCin)$ as a function of observation time
    $T$.  Here $\Izz\MCin = \SI{2e38}{kg.m^2}$, $n\MCin = 4$,
    $f\MCin = \SI{1000}{Hz}$, $\fdot\MCin = \SI{-1e-9}{Hz.s^{-1}}$, and
    $\cD = \SI{30}{Hz^{-1/2}}$, which implies $h_0 = \num{8.1e-25}$,
    $\epsilon = \num{3.8e-7}$, and $\pmp = \SI{2.3e19}{T.m^3}$.
    The input values (dashed lines) are plotted against the median, 16th and 84th
    percentiles for $10^6$ samples.
  }
  \label{fig:single_NS}
\end{figure}

Figure~\ref{fig:single_NS} illustrates how the errors in the inferred neutron
star properties scale with observation time. Here the inputs are
fixed to the representative values $\Izz\MCin = \SI{2e38}{kg.m^2}$, $\epsilon\MCin = \num{1.2e-7}$,
$\pmp = \SI{7.2e18}{T.m^3}$, and output values
$(\Izz\MCout, \epsilon\MCout, \pmp\MCout)$ are simulated for different $T$,
assuming a sensitivity depth $\cD = \SI{30}{Hz^{-1/2}}$.  As expected, the errors
of the inferred parameters decrease with increasing observation time. For
$T \gtrsim \SI{2}{years}$ the errors in $\Izz$, $\epsilon$, and $\pmp$
asymptote to the error due to $\dist$, consistent with
Eqs.~\eqref{eq:Izz-rel-err-dea}~--~\eqref{eq:mp-rel-err-dea}.  We neglect the
possibility that the error in distance may be improved over time if better models
of the galactic electron density distribution become available.

\begin{figure}
  \centering
  \includegraphics[width=\columnwidth]{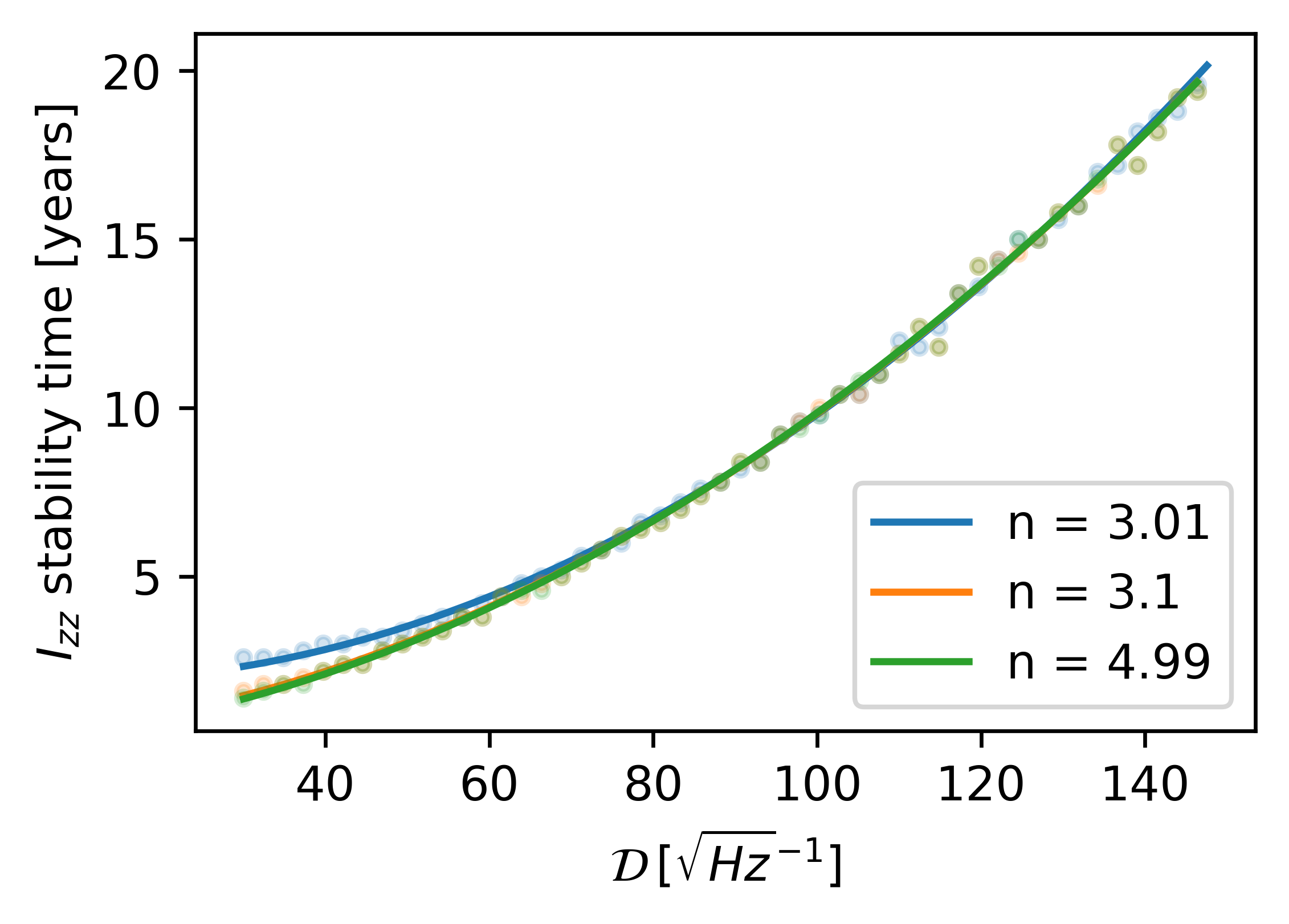}
  \caption{
    $\Izz$ stability time versus sensitivity depth $\cD$ for braking indices
    $n\MCin = 3.01, 3.1, 4.99$.  Here $\Izz\MCin = \SI{2e38}{kg.m^2}$,
    $f\MCin = \SI{1000}{Hz}$, $\fdot\MCin = \SI{-1e-9}{Hz.s^{-1}}$, and $T = 1$~year,
    with values for $h_0$, $\epsilon$, and $\pmp$
    implied by Eqs.~\eqref{eq:h0},~\eqref{eq:epsilon}, and~\eqref{eq:mp}
    respectively.  Plotted are a subsampling of the results from $10^6$ samples
    (light-coloured dots) and best-fit curves (dark-coloured lines).
  }
  \label{fig:sensitivity_depth}
\end{figure}

We define the ``stability time'' for $\Izz$, $\epsilon$, and $\pmp$ as the time
required for the normalised relative errors $\NRE$ of each property to reach
$1.1$, i.e. to within 10\% of the asymptotic distance error [see
Eq.~\eqref{eq:norm-median-relative-error}].  Figure \ref{fig:sensitivity_depth}
plots the stability time for $\Izz$ as a function of $n$ and $\cD$ for signals
with $\Izz\MCin = \SI{2e38}{kg.m^2}$, $f\MCin = \SI{1000}{Hz}$, and
$\fdot\MCin = \SI{-1e-9}{Hz.s^{-1}}$. We see that, for continuous wave signals
at $\cD \sim \SI{30}{Hz^{-1/2}}$ initially detected in an all-sky survey, the
asymptotic error in $\Izz$ is approached after a few years observing with a
fully-coherent follow-up search, which would include analysing both archival and
future data.  For continuous waves detected from known pulsars, where
$\cD \approx \SI{150}{Hz^{-1/2}}$, the asymptotic errors in $I_{zz}$ are not approached
until the star is observed for $T\approx \SI{20}{years}$ which is an unrealistic time
span to consider. Note, however, that the
definition of ``stability time'' here assumes the detector sensitivity $S_h$
remains constant; in reality $S_h$ is likely to decrease over
time~\citep{KAGREtAl2020:PrObLclGrvTrALAVK}, particularly if third-generation
gravitational wave detectors are constructed~\citep{BailEtAl2021:GrvPhAst2020}.
Such improvements would decrease the sensitivity depth $\mathcal{D}$ of a detected
signal such that the inferred parameters would converge to the asymptotic distance 
error faster than suggested in figure \ref{fig:sensitivity_depth}.

\begin{figure}
  \centering
  \includegraphics[width=\columnwidth]{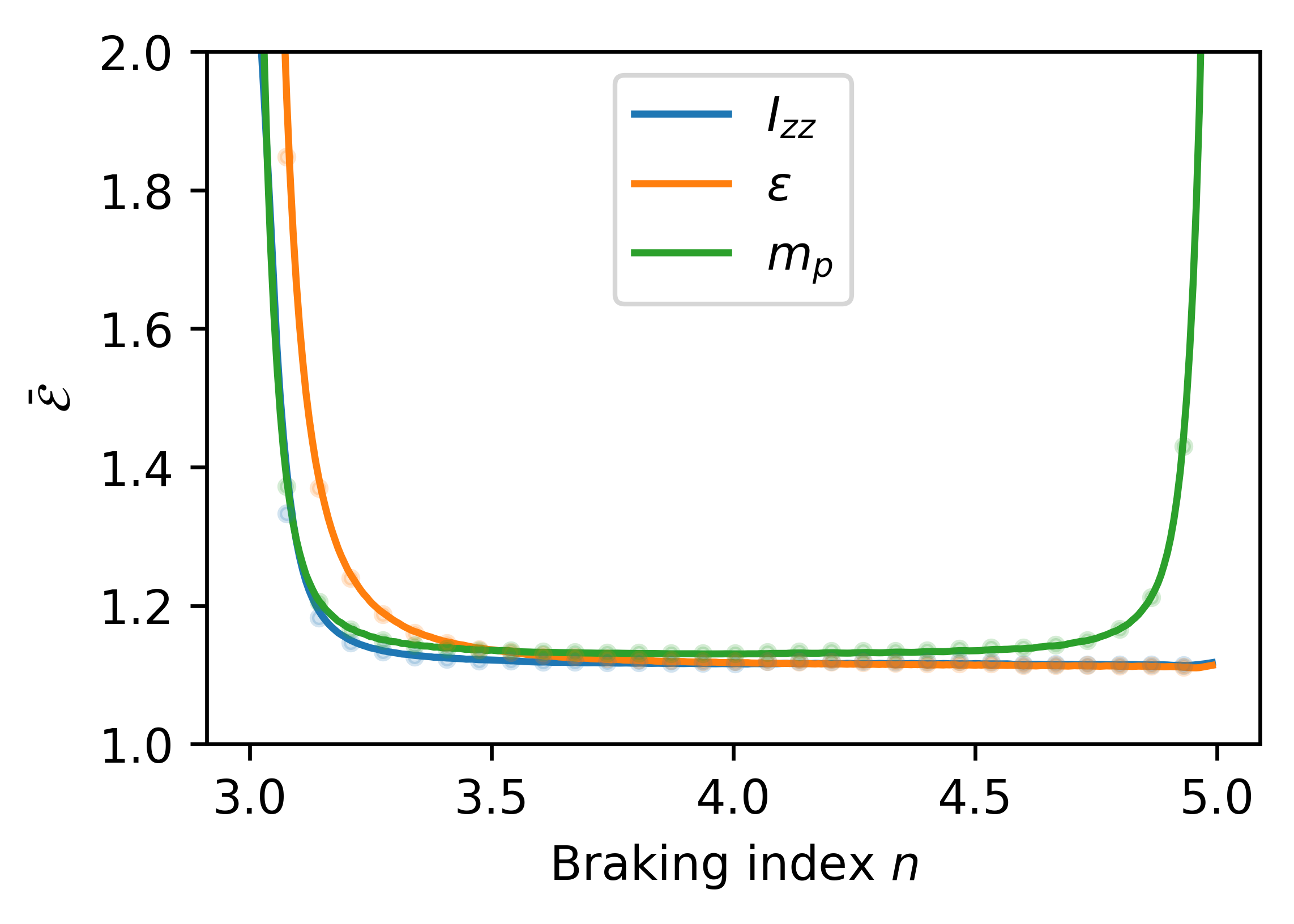}
  \caption{
    Normalised relative errors ($\NRE$) as a function of braking index $n\MCin$.  Here
    $\Izz\MCin = \SI{2e38}{kg.m^2}$, $f\MCin = \SI{1000}{Hz}$, $\fdot\MCin =
    \SI{-1e-9}{Hz.s^{-1}}$, $T = 1$~year, and $\cD = \SI{30}{Hz^{-1/2}}$, with
    values for $h_0$, $\epsilon$, and $\pmp$ implied by
    Eqs.~\eqref{eq:h0},~\eqref{eq:epsilon}, and~\eqref{eq:mp} respectively.
    Plotted are a subsampling of the results from $10^6$ samples
    (light-coloured dots) and best-fit curves (dark-coloured lines).
  }
  \label{fig:braking_index}
\end{figure}

Figure~\ref{fig:braking_index} plots normalised relative errors in $\Izz$,
$\epsilon$, and $\pmp$ as functions of $n$ for signals with
$\Izz\MCin = \SI{2e38}{kg.m^2}$, $f\MCin = \SI{1000}{Hz}$,
$\fdot\MCin = \SI{-1e-9}{Hz.s^{-1}}$, and $\cD = \SI{30}{Hz^{-1/2}}$.  The
neutron star properties $\Izz$ and $\epsilon$ are best estimated where the
star is losing almost all energy in gravitational waves ($n \approx 5$), as
expected. On the other hand, the electromagnetic property $\pmp$ is best
estimated where the star is losing its energy through both electromagnetic and
gravitational radiation ($n \approx 4$).  When energy loss through electromagnetic
radiation is more dominant ($n \approx 3$), the errors for all three properties
are larger than the $n\approx 4$ case. This is consistent with
Eq.~\eqref{eq:Izz-rel-err-dea} - \eqref{eq:mp-rel-err-dea} and is because the continuous wave observation cannot measure the spindown parameters accurately when the neutron star only weakly emits continuous waves. However, note that these results
are based on the techniques described in section \ref{sec:framework}. It may
be possible for electromagnetic astronomers to use alternate techniques to measure $\pmp$
with lower errors for neutron stars with certain braking indices.

\begin{figure*}
  \includegraphics[width=\textwidth]{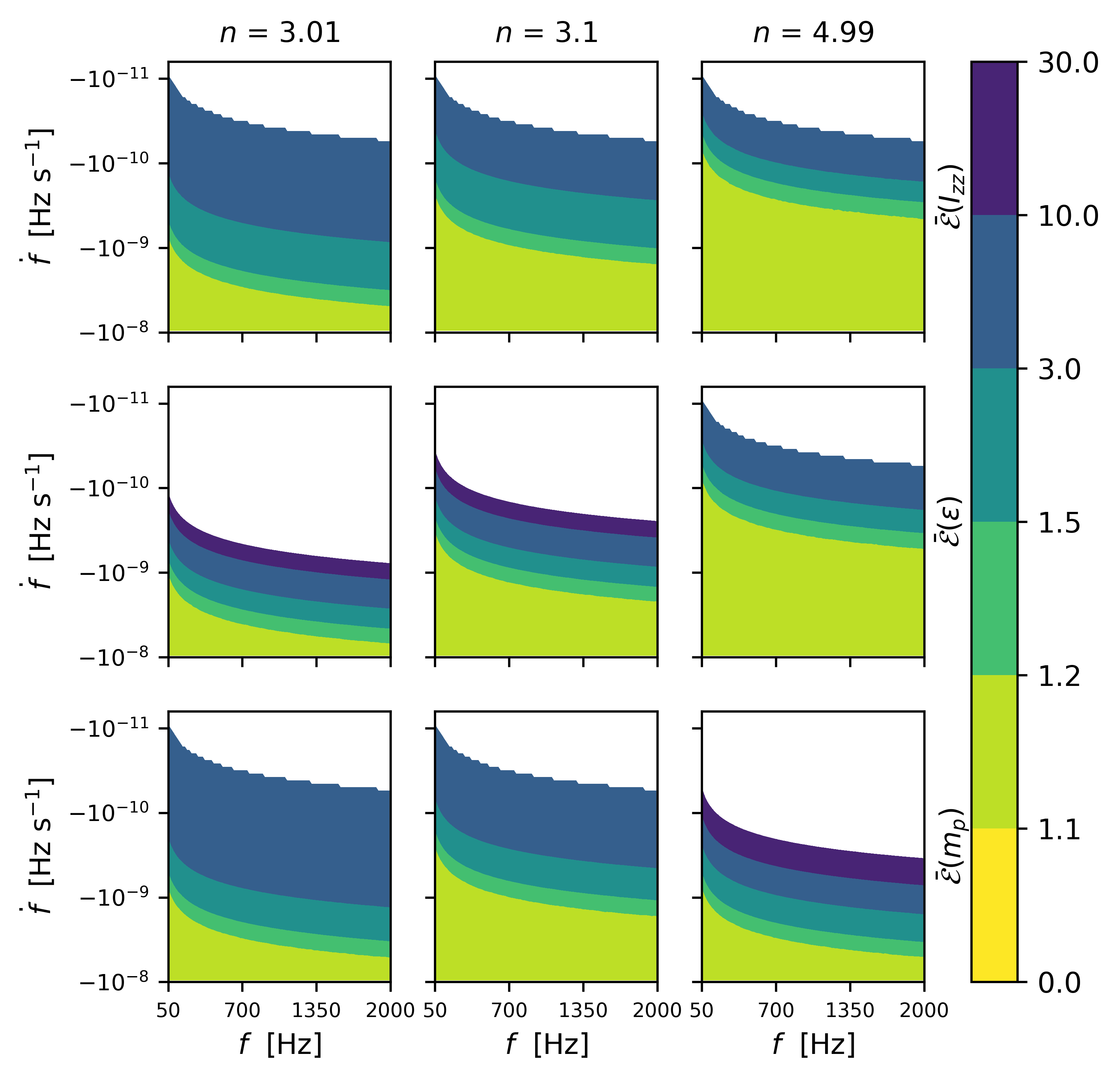}
  \caption{
    Normalised relative errors (top to bottom rows) $\NRE(\Izz)$, $\NRE(\epsilon)$,
    $\NRE(\pmp)$, for braking indices (left to right columns) $n = 3.01$, $3.1$,
    $4.99$, as functions of $f$ and $\fdot$.  Plotted are the median
    errors over $\Izz$ and $\xi$, and for $T = 1$~year and $\cD = \SI{30}{Hz^{-1/2}}$.
    The different colours represent the values of the normalised relative error and the white areas indicate where $\NRE \ge 30.0$ and/or where $3 < n\MCout < 5$
    no longer holds. A total of $10^6$ samples were used in each plot.
  }
  \label{fig:heatmap}
\end{figure*}

Figure~\ref{fig:heatmap} plots normalised relative errors $\NRE(\Izz)$,
$\NRE(\epsilon)$, and $\NRE(\pmp)$ as functions of $n$, $f$, and $\fdot$, taking
the median errors over the sampled ranges of $\Izz$ and $\xi$
given in Section~\ref{sec:MC_params}.  We assume $T = 1$~year and
$\cD = \SI{30}{Hz^{-1/2}}$, which is relevant to a continuous wave signal
detected in an all-sky continuous wave survey.  The errors in all three neutron
star properties are smallest at the highest spin-down rates
($\fdot \approx \SI{-e-8}{Hz.s^{-1}}$), where rate of rotational kinetic energy
loss from the star is highest, and lowest frequencies ($f\approx \SI{50}{Hz}$). Once $|\fdot| \lessapprox \SI{e-11}{Hz.s^{-1}}$,
the errors are sufficiently large that $n$ cannot be reliably measured, the
restriction $3 < n\MCout < 5$ is no longer satisfied, and most Monte Carlo samples
must be discarded (see Section~\ref{sec:comp-outp-param}).
For each heatmap, the error as a function of $|\fdot|$ increases more rapidly for lower $f$ than for higher $f$, consistent with the $f^{-2}$ dependence of the $T^{-3}$ terms in Eqs.~\eqref{eq:Izz-rel-err-dea}~--~\eqref{eq:mp-rel-err-dea}.

Figure~\ref{fig:heatmap} suggests that normalised relative errors
$\NRE \lesssim 1.2$ are achievable over much of the $f$--$\fdot$ parameter space
typically searched over for continuous waves, and particularly for rapidly
spinning-down sources ($|\fdot| \gtrapprox \SI{e-9}{Hz.s^{-1}}$). This implies
errors in $\Izz$ of $\sim 32\%$, and errors in $\epsilon$ and $\pmp$ of
$\sim 16\%$. Given that models of non-axisymmetrically deformed neutron
stars~\citep{BonaGour1996:GrvWPlEmMgFIDst, UshoEtAl2000:DfrAcNtSCGrvWEm,
  Cutl2002:GrvWvNtSLTrdBFl, Owen2005:MxElDfrCmSEEqtS,
  PaynMela2006:FrSGrvRGHydOsMgCMANS, HaskEtAl2008:MdlMgnDfNtSt,
  VigeMela2009:IEDtcGrvRdMgCMAcNS, WettEtAl2010:SnMgnCnMnAccNS,
  PriyEtAl2011:QdrMMgnCnMnAcNSEES} typically predict $\epsilon$ only to an order
of magnitude, the error in $\epsilon$ should be sufficient to test such models.
A $\sim 30\%$ error in $\Izz$ is of similar magnitude to measurements of $\Izz$
for PSR~J0737$-$3039A. \citet{MiaoEtAl2022:MmInPJ07LIGNI} found errors of
$\sim 10$~--~$20\%$ after assuming an equation of state; without that assumption
the errors in $\Izz$ increase by a factor of $\sim 4$.  In comparison, no explicit
assumptions regarding the neutron star equation of state are required for the
framework of Section~\ref{sec:framework} or the results presented in Section
\ref{sec:results}. Estimates of $\pmp$ using this framework could also be compared to those
estimates for known pulsars and serve as an independent verification of such measurements.

\begin{table}
    \centering
    \begin{tabular}{|c|c|c|}
    \hline 
    Pulsar & Crab (J0534+2200) & Vela (J0835-4510) \\
    \hline
    $f_{\text{GW}}$ (\si{Hz}) & 59.2 & 22.4 \\
    $\dot{f}_{\text{GW}}$ (\si{Hz.s^{-1}}) & \num{-7.4e-10} & \num{-5.6e-11} \\
    $r$ (\si{kpc}) & 2.0 & 0.157 \\
    $\Delta r/r$ & 0.25 & 0.066 \\
    $h_0^{95\%}$ & \num{1.5e-26} & \num{2.2e-25} \\
    \hline
    \end{tabular}
    \caption{Properties of two young pulsars. Here $h_0^{95\%}$ is the 95\% confidence level upper limits on the gravitational wave amplitude measured by \citet{Abbott2020}. Adapted from \citet{Abbott2020}.}
    \label{table: data}
\end{table}
\begin{table}
    \centering
    \begin{tabular}{|c|c|c|}
    \hline 
    Pulsar & Crab (J0534+2200) & Vela (J0835-4510) \\
    \hline
    $n^{\text{max}}$ & 3.0025 & 3.0050 \\
    $\RE(\Izz)$ & 0.97 & 0.99 \\
    $\RE(\epsilon)$ & 29 & 93 \\
    $\RE(\pmp)$ & 0.82 & 0.91 \\
    \hline
    \end{tabular}
    \caption{Maximum braking index ($n^{\text{max}}$) and relative errors ($\RE$) for two young pulsars. Here $n^{\text{max}}$ is computed by rearranging Eq. \ref{eq:Izz} and substituting in the known $f$ and $\dot{f}$, measured $h_0^{95\%}$, and assuming $\Izz = \SI{1e38}{kg.m^2}$ (which is consistent with \citealt{Abbott2020}). The relative errors are computed for $T = 1$~year, $\cD = \SI{150}{Hz^{-1/2}}$, and for $10^6$ samples.}
    \label{table: results}
\end{table}

Table~\ref{table: data} shows the properties of two young pulsars. By rearranging Eq.~\ref{eq:Izz} for $n$, it is possible to compute the maximum braking indices ($n^{\text{max}}$) consistent with the observational upper limits $h_0^{95\%}$. This is computed in Table~\ref{table: results}, where it is necessary to assume $\Izz=\SI{1e38}{kg.m^2}$ which is consistent with \citealt{Abbott2020}. If continuous waves from the two pulsars were observed with $n = n^{\text{max}}$, Table~\ref{table: results} also shows the relative errors of the parameters that would be inferred. These relative errors are larger than those in Figure~\ref{fig:heatmap} because of the extremely low braking indices and increased sensitivity depths for continuous waves from known pulsars. Nevertheless, this shows how continuous waves could infer properties of known pulsars that are otherwise difficult to measure.

\section{Assumptions} \label{sec:assumptions}

This section elaborates on some of the assumptions made in this paper.

We assume that continuous waves will eventually be detectable by contemporary
and/or future gravitational wave detectors.  This remains uncertain. The lowest
bounds on $\epsilon \propto 10^{-11}$ from magnetic field
distortions~\citep{BonaGour1996:GrvWPlEmMgFIDst} are small enough that only a few detections, at best, may be expected in the next generation of detectors~\citep{Pitk2011:PrObCntGrvWKPl}.
Stars with stronger magnetic fields ($B \sim 10^{15}$~Gauss) lead to larger ellipticities
$\epsilon \gtrsim 10^{-6}$~\citep{Cutl2002:GrvWvNtSLTrdBFl,
HaskEtAl2008:MdlMgnDfNtSt} which are more likely detectable by the current generation of gravitational wave detectors. It is also possible that the internal magnetic fields of neutron stars could be stronger than their surface fields \citep{Lasky2015, Bransgrove2017}. For known pulsars, only a small fraction are
likely to be detectable, particularly if the fraction of rotational kinetic
energy emitted in gravitational waves is small~\citep{Pitk2011:PrObCntGrvWKPl}.
That said, the $\cO(10^3)$ known pulsars may not be representative of the
$\cO(10^8)$ population of galactic neutron stars~\citep{Palomba2005,
  Knispel2008, Wade2012, Cieslar2021, ReedEtAl2021:MdGlNSPplUCnGrvS}, which could include a sub-population of
strong gravitational wave emitters or ``gravitars''.

We assume that Eq.~\eqref{eq:energy_balance} is a reasonable starting point
for modelling the energy radiated by neutron stars. It is generally assumed that
electromagnetic radiation from known pulsars is predominately dipolar, that neutron stars are triaxial rotors, and that continuous wave radiation would be predominately
quadrupolar~\citep{OstrGunn1969:NtrPlsITh}. Recent measurements of hot surface regions of several neutron stars using NICER provide evidence for non-dipolar magnetic fields in millisecond pulsars \citep{Bilous2019, Riley2019, Riley2021}. However, alternate models with multipolar magnetic fields are not well understood \citep{Gralla2017, Lockhart2019, Riley2019} and their formation during stellar evolution is also unclear \citep{Mitchell2015, Gourgouliatos2018}. While dipolar magnetic fields are a reasonable starting point for this analysis, future work could extend the framework developed here to more complex magnetic field configurations.

The energy emission assumptions of Eq.~\eqref{eq:energy_balance} predicts $3 < n < 5$, which is at odds with measured braking indices from radio pulsars which span orders of magnitude outside this
range~\citep{JohnGall1999:PlBrkInRvs, ZhanXie2012:WDBrkInPlSRnM100Mll, Lower2021}.  Modified models for
pulsar emission have been proposed to explain the observed braking
indices~\citep{AlleHorv1997:ImpCnsObBrIYPSp, Mela1997:SpObRtCrrOMgn,
  XuQiao2001:PlBrkInTEmsMd, AlvaCarr2004:MnpPlSpn,
  YueEtAl2007:WCBrkIndTUAbNtPln, HamiEtAl2015:BrkInIslPl} including the addition
of gravitational waves \citep{deArEtAl2016:GrvWvPlMsrBrI,
  ChisEtAl2018:AnlAppStPlSp}.  On the other hand, accurate phase-connected measurement of a
second time derivative of the rotation frequency needed to compute $n$ is
challenging~\citep[cf.][]{JohnGall1999:PlBrkInRvs}. Existing measurements
of $n$ are generally dominated by timing noise~\citep{HobbEtAl2004:LngTmObs374Pl,
  HobbEtAl2010:AnlTmIrr366Pl}, with some possible exceptions~\citep{ArchEtAl2016:HgBrkInPls,LaskEtAl2017:BrInMllMgn}.

Prospects for an accurate determination of $n$ may be improved by a continuous
wave detection. Since gravitational wave detectors are omni-directional,
gravitational wave data is recorded at a much higher duty cycle
($\gtrsim 70\%$; \citealt{KAGREtAl2020:PrObLclGrvTrALAVK}) than typical pulsar
observing cadences (e.g. $\lesssim 35 \text{hours/year} \sim 0.4\%$;
\citealt{Lam2018:OpPTAObsCSnLwGrvS}). Although $\fddot$ would not be
resolved in all-sky continuous wave
surveys, which sacrifice phase resolution in favour of reduced computational
cost, a candidate from such a survey would then be followed up using a fully
phase-coherent search in a restricted parameter space around the candidate. Such
a search would be computationally inexpensive, and would be able to resolve
$\fddot$ to a resolution $\sim \cD / T^{7/2}$ [cf. Eq.~\eqref{eq:spindown_cov}].

Pulse emission from radio pulsars is subject to various noise
sources~\citep{deKoAnze1993:SmAnlPrNBnXPls, ArchEtAl2008:RNsAnmXPlTmRsd,
  LentEtAl2016:SNSysStPrFIntPTADRl, GoncEtAl2021:IdnMtgNSrPrPlTDS}.  Individual
pulses from radio pulsars are highly variable, and achieve a stable pulse
profile once averaged over many cycles~\citep{Kram2005:Pls}.  It remains
to be seen whether detected continuous wave signals will suffer from comparable
noise sources~\citep{AshtEtAl2015:ETNTrNrrChSCnGrvWP, Suvorova2016, Myers2021a, Myers2021b}.  Gravitational waves,
being weakly interacting, are not perturbed by matter along the line of sight to
the star, unless the signal is lensed~\citep{BiesHari2021:GrvLnCnGrvWv}.
Furthermore, unlike electromagnetic emission that arises from the outer surface and plasma of the
star, where a small fraction of the neutron star mass is located, gravitational wave emission
arises from the rotating mass quadrupole.  Physical processes
within the star would need considerable energy to perturb the star's
rotation, and hence the continuous wave signal, in order to achieve a level
of noisiness comparable to timing noise observed in radio pulsars.
Superfluid vortices within the star's interior are suspected of being
responsible for glitches~\citep{EysdMela2008:GrvRdtPlGl,
  WarsMela2011:GrsMdPlGlt, HoEtAl2015:PnSprMsrMsUPGl,
  HaskEtAl2020:TrPnSprNtSPGRcv, LIGOEtAl2021:CnsLODGrvEmDRGlPPJ0} which do
perturb the star's rotation and may affect the detectability of continuous
waves~\citep{AshtEtAl2018:SmcGlCntSrMt}.  Glitches, however, are observed as
discrete events even in prolifically glitchy
pulsars~\citep{HoEtAl2020:RtBGltNTmGlPJ05} and the extent to which they could constitute a
persistent noise source in detected continuous wave signals is unknown \citep{Yim2022}.  Should continuous
waves measure a braking index from $n \notin [3, 5]$, this might represent stronger
evidence for new physics than current radio pulsar observations.

Finally, we assume that the neutron star also emits electromagnetic radiation,
and that a measurement of its distance can be obtained. Neutron stars are
expected to possess magnetic fields~\citep{Reis2001:MgnFlNtStOvr} and will
therefore (provided that the field is not symmetric about the star's rotation
axis) emit electromagnetic radiation. Continuous waves may first be detected either
from a known pulsar, or as a gravitational-wave-only candidate from an all-sky
survey; in either case, observations over $T \gtrsim 1$~year would give the sky
position of the source to sub-arcsecond resolution~\citep{Riles2013, Riles2017}. This would
facilitate further electromagnetic observations to either detect an
electromagnetic counterpart, or else refine the properties of one already known.
Other methods exist to measure stellar distances in the absence of a radio
pulsar detection; parallax may be used to determine the distances to nearby
neutron stars~\citep{Seto2005:GrvWAstRRtNSEsTDs,WaltEtAl2010:RvsPrIsNtSRJ18UHI}, while distances to neutron
stars in supernova remnants may be inferred through observation of the radial
velocities of the surrounding ejecta~\citep{ReedEtAl1995:ThrStCssSpRISpS}. These
methods yield comparable uncertainties to radio pulsar distances.

\section{Summary} \label{sec:summary}

This paper presents a first analysis of what properties may be inferred from a
neutron star radiating both electromagnetic and detectable continuous
gravitational waves. We develop a simple Fisher information-based parameter estimation framework, which gives estimates of the uncertainties for the stellar moment of inertia $\Izz$, equatorial ellipticity $\epsilon$, and component of the magnetic dipole moment perpendicular to its rotation axis $\pmp$. This framework does not assume a particular neutron star equation of state and only requires a
detection of continuous waves and a measurable distance to the star.

Monte Carlo simulations over a parameter space of gravitational wave frequency
and its derivatives, typical of that covered by all-sky continuous wave surveys,
demonstrate that the relative errors in $\Izz$, $\epsilon$, and $\pmp$
asymptote to 14--27\%, assuming a 20\% error in distance. The observation
time required to reach these limits may be as little as a few years for a strong
continuous wave signal detected in an all-sky survey; for weaker signals, such
as those potentially associated with known pulsars, longer observations may be required.
We also find that the errors of the inferred parameters tend to be smaller when the braking index is close to $n\approx4$, when $f$ is smaller and when $|\dot{f}|$ is larger.

Future work could extend the assumed neutron star energy loss model of
Eq.~\eqref{eq:energy_balance} to include a more complex model of the neutron
star magnetic field, e.g.~\citet{LaskMela2013:TlTMgFNtSTGrvWSgn}. Recasting
the parameter inference in a Bayesian framework would also be advantageous as it would avoid the coordinate singularities present in the Fisher matrix approach and make use of prior information from other
gravitational wave and electromagnetic observations of neutron stars.


\section*{Acknowledgements}

We thank Lucy Strang, Lilli Sun, Matthew Bailes, and Ryan Shannon for helpful
discussions.  This research is supported by the Australian Research Council
Centre of Excellence for Gravitational Wave Discovery (OzGrav) through project
number CE170100004.

\section*{Data Availability}

The data underlying this article will be shared on reasonable request to the
corresponding author(s).


\bibliographystyle{mnras}
\bibliography{Bibliography}


\appendix


\bsp	
\label{lastpage}
\end{document}